\documentclass{aa}
\usepackage{graphics}
\usepackage{txfonts}
\usepackage{natbib}

\newcommand{\kms}   {km~s$^{-1}$}

\def\gs{\mathrel{\raise0.35ex\hbox{$\scriptstyle >$}\kern-0.6em
\lower0.40ex\hbox{{$\scriptstyle \sim$}}}}
\def\ls{\mathrel{\raise0.35ex\hbox{$\scriptstyle <$}\kern-0.6em
\lower0.40ex\hbox{{$\scriptstyle \sim$}}}}

\begin{document}

   \title{Improved VLBI astrometry of OH maser stars}
   \titlerunning{Improved VLBI astrometry of OH maser stars}


   \author{W.H.T. Vlemmings\inst{1}\and
        H.J. van Langevelde\inst{2,3}
          }

   \offprints{WV (wouter@astro.uni-bonn.de)}

   \institute{Argelander Institute for Astronomy, University of Bonn,
     Auf dem H{\"u}gel 71, 53121 Bonn, Germany
        \and
              Joint Institute for VLBI in Europe, Postbus 2, 
                7990~AA Dwingeloo, The Netherlands
	\and     
        Sterrewacht Leiden, Postbus 9513, 2300 RA Leiden, 
              The Netherlands}

   \date{Received ; accepted }


\abstract{}{Accurate distances to evolved stars with high mass loss
  rates are needed for studies of many of their fundamental
  properties. However, as these stars are heavily obscured and
  variable, optical and infrared astrometry is unable to provide
  enough accuracy.}{Astrometry using masers in the circumstellar
  envelopes can be used to overcome this problem. We have observed the
  OH masers of a number of Asymptotic Giant Branch (AGB) stars for
  approximately 1~year with the Very Long Baseline Array (VLBA). We
  have used the technique of phase referencing with in-beam
  calibrators to test the improvements this technique can provide to Very
  Long Baseline Interferometry (VLBI) OH maser astrometric
  observations.}{We have significantly improved the parallax and
  proper motion measurements of the Mira variable stars U~Her, S~CrB
  and RR~Aql. }{It is shown that both in-beam phase-referencing
  and a decrease in solar activity during the observations
  significantly improves the accuracy of the astrometric
  observations. The improved distances to S~CrB ($418^{+21}_{-18}$~pc)
  and RR~Aql ($633^{+214}_{-128}$~pc) are fully consistent with
  published P--L relations, but the distance to U~Her
  ($266^{+32}_{-28}$~pc) is significantly smaller. We conclude that
  for sources that are bright and have a nearby in-beam calibrator,
  VLBI OH maser astrometry can be used to determine distances to OH
  masing stars of up to $\sim2$~kpc. }

   \maketitle

\section{Introduction}

VLBI astrometric observations of circumstellar OH masers can be used
to determine the proper motion and parallax of enshrouded maser
bearing stars \citep[][, hereafter vL00 and V03]{vL00,
  Vlemmings03}. Obtaining accurate distances to these stars allows the
inclusion of the more extreme Mira stars in studies of the fundamental
properties of these stars, like the pulsation and mass-loss
mechanism. Accurate parallaxes also allow the growing number of
optical and infrared interferometric observations of AGB stars to be
used to determine actual stellar sizes \citep[e.g.][]{Menneson02,
  Wittkowski05}. Currently, most investigations of pulsating stars are
dependent on Hipparcos \citep{Perryman97}, which is biased against the
highly obscured stars with high mass loss. Unfortunately, as these
stars are variable, the Hipparcos distances are often very
uncertain. Maser astrometry can remedy this situation. In order to use
the maser positions to monitor the stellar trajectory, an assumption
has to be made about the motion of the masers with respect to the
star. In V03, it was shown that for several stars the brightest, most
blue-shifted circumstellar maser spot corresponds to the {\it
  Amplified Stellar Image}, as was previously hypothesized by
\cite{Norris84} and \cite{Sivagnanam90}.  However, not all stars show
such a maser spot, depending on for instance an inhomogeneous
distribution of the masing gas \citep{Vlemmings02PhD}. Still the
observations in V03 indicate that even without an amplified stellar
image, VLBI astrometry of OH
masers yields highly accurate parallax and proper motion results. The
only assumption needed for this is that these spots have linear
motions in the shell.

The technique of VLBI astrometry has undergone extensive developments
in the past several years. The feasibility of annual parallax
measurements has now been proven at several different frequencies. At
frequencies around 1.6~GHz, current astrometric observations provide
pulsar velocities and distances out to several kpc
\citep[e.g.][]{Brisken02, Chatterjee05}. At higher frequencies, H$_2$O
and methanol maser astrometry is being used to obtain star-forming
region distances up to several kpc with close to $\sim 2\%$ accuracy
\citep[e.g.][]{Hachisuka06, Xu06}. H$_2$O maser observations have also
recently been used to measure the parallax of the Mira star UX~Cyg
\citep{Kurayama05} and additional H$_2$O and SiO maser AGB astrometric
observations are expected from the VERA project \citep{Honma03}.

In this paper we present a follow-up on our earlier successful VLBI OH
maser astrometry campaign (vL00, V03). In \S~\ref{obs} we discuss the
observations, data reduction and error analysis. In \S~\ref{res} we
give the results of our parallax and proper motion determination. The
results are compared with previous distance measurements in
\S~\ref{disc} along with a discussion about the internal maser motions
in U~Her. The conclusions are presented in \S~\ref{concl}.

\section{Observations \& Data reduction}
\label{obs}

\begin{figure*}[ht!]
  \begin{center}
   \resizebox{\hsize}{!}{\includegraphics{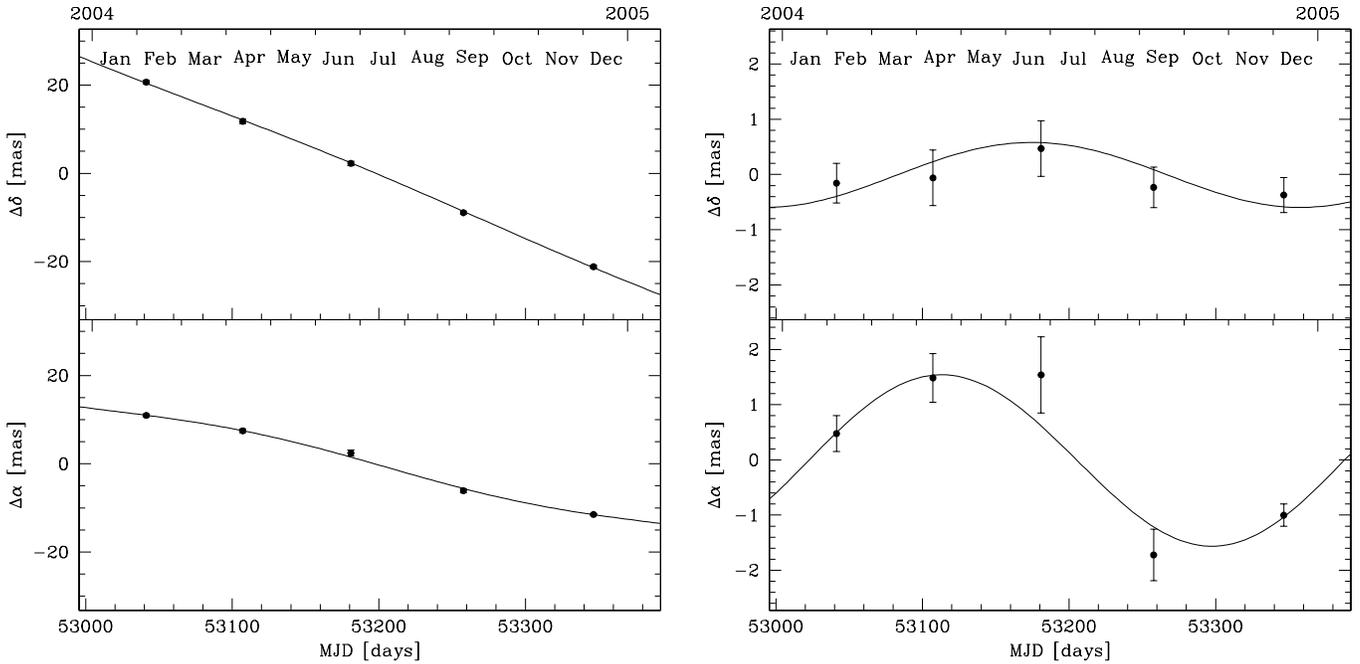}}
  \end{center}
   \hfill \caption{The position of the brightest 1665 MHz maser spot
     of RR~Aql with respect to the in-beam calibrator (NVSS
     J195655-013615). The error bars on the positions indicate the
     position fitting errors described in \S~\ref{error}. (left) The
     solid line is the best fitting parallax and proper motion
     trajectory for the 5 epochs with usable observations. (right)
     The best fitted parallax signature after subtracting the fitted
     proper motions.}
   \label{fig1.rraql}
\end{figure*}

\begin{figure*}
   \resizebox{\hsize}{!}{\includegraphics{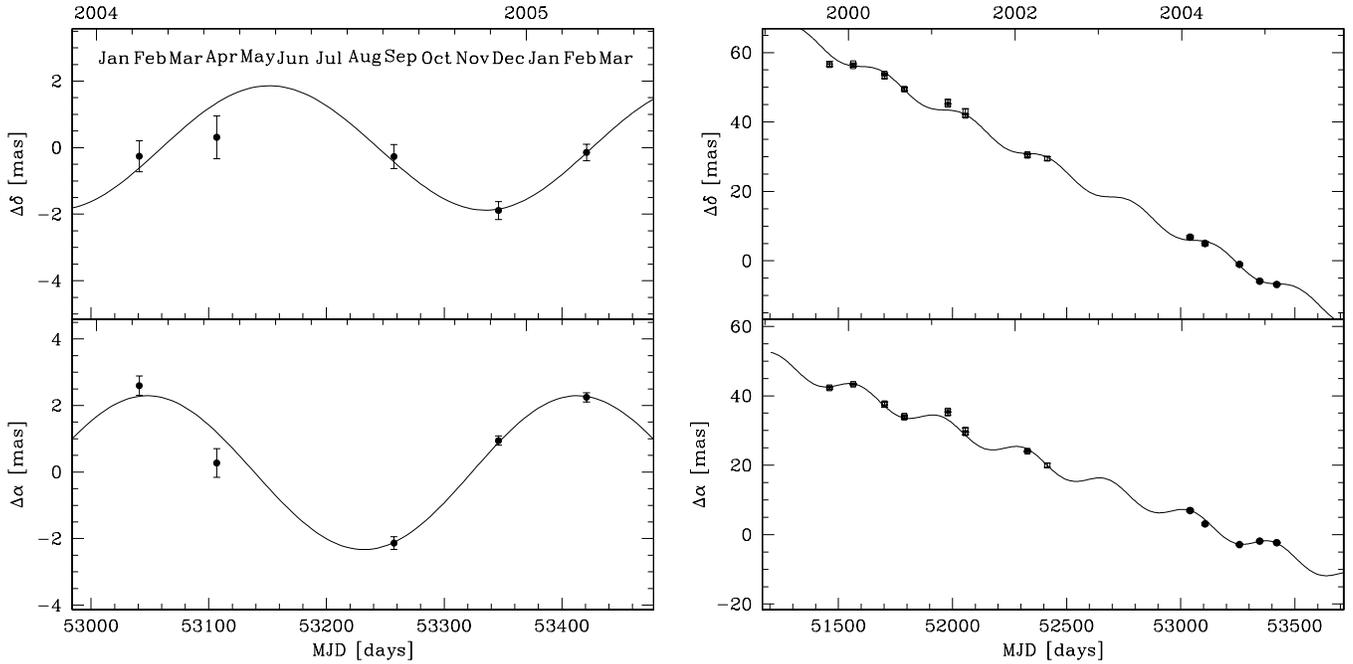}}
   \hfill \caption{Similar to Fig.~\ref{fig1.rraql}. The position of
     the brightest 1667 MHz maser spots of S~CrB with respect to
     J1522+3144, which was used as in-beam calibrator during the 5
     epochs presented here. Drawn is the best fitting parallax signature after subtracting the proper motions for the 5 new epochs with in-beam
     calibration (left) and the proper motion and parallax trajectory including the previous observations of the
     brightest 1665 and 1665 MHz OH maser spots (V03) for a total of
     13 epochs (right).}
   \label{fig2.scrb}
\end{figure*}

\begin{figure*}
   \resizebox{\hsize}{!}{\includegraphics{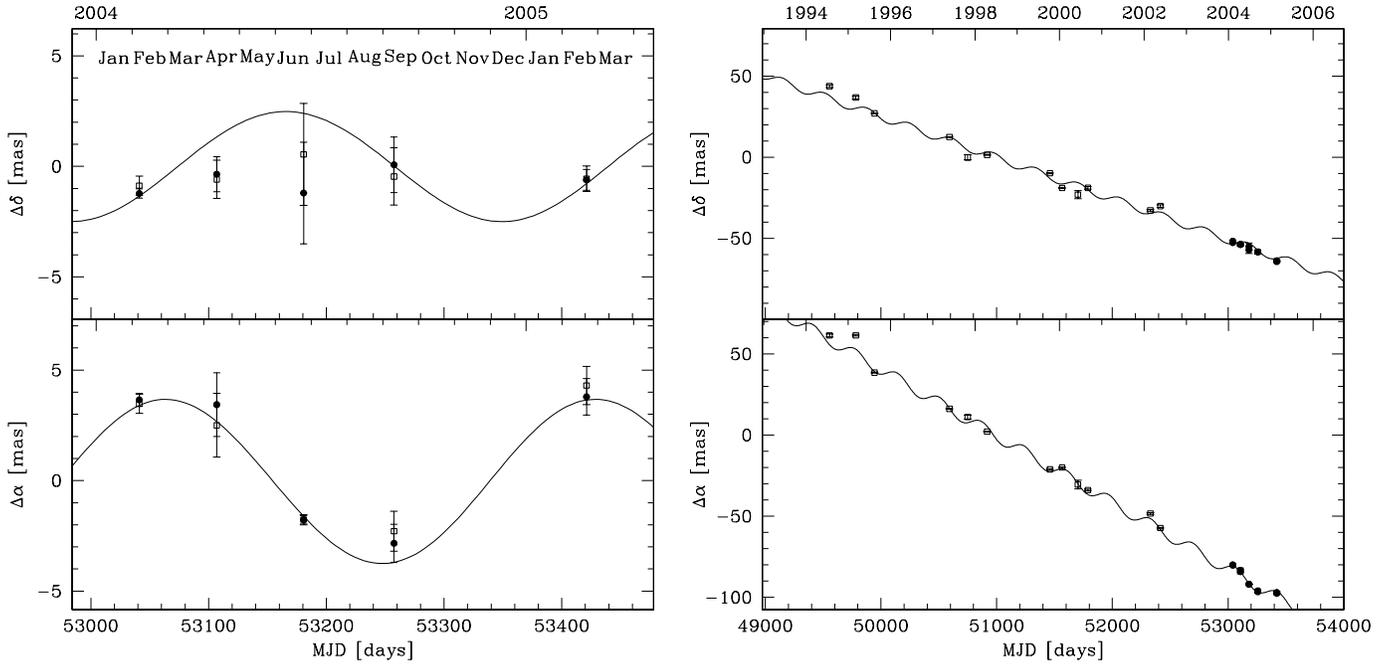}}
   \hfill \caption{ Similar to Fig.~\ref{fig1.rraql}. The position of
     the most blue-shifted 1665 and 1667 MHz maser spots of U~Her,
     that have been found to correspond to the stellar image, with
     respect to the primary calibrator J1630+2131. Drawn is the best
     fitting parallax signature after subtracting the proper
       motions for combining the results on both maser transitions for
       the 5 recent epochs of observations (left), and the proper
       motion and parallax trajectory combining the new observations
       with the 12 previous epochs (right).}
   \label{fig3.uher}
\end{figure*}

\begin{table*}
\caption{The sample}
\begin{tabular}{|l||c|c|c|rr|c|}
\hline
{ Source} & { Period} & $V_{\rm star} $ & Hipparcos Parallax &
\multicolumn{2}{c|}{{ Hipparcos Proper motion}} & $d_{P-L}$$^b$\\
& { (days)} & { (km~s$^{-1}$)} & (mas) &
    \multicolumn{2}{c|} {{ RA, dec (mas/yr)}} & (pc)\\
\hline
\hline
S~CrB & 360 & 0.0 & $2.40 \pm 1.17$$^a$ & $-8.33 \pm 0.93$ & $-11.55$$\pm$$0.62$& 470 \\
U~Her & 406 & -14.5 & $1.88 \pm 1.31$$^a$ & $-16.84 \pm 0.82$ & $-9.83$$\pm$$0.92$ & 380 \\
RR~Aql & 394 & 27.8 & $2.48 \pm 2.57$ & $-24.01 \pm 4.18$ & $-47.66$$\pm$$2.80$ & 540 \\
\hline
\multicolumn{7}{l}{$^a$ recalculated by \cite{Knapp03}}\\
\multicolumn{7}{l}{$^b$ from \cite{Whitelock00}}\\
\end{tabular}
\label{sam1}
\end{table*}

The positions of the 1.6~GHz circumstellar OH masers of a sample of 6
AGB stars were monitored at 6 epochs spaced by 2-3~months between
February 2004 and April 2005. The observations were perfomed using the
NRAO\footnote{The National Radio Astronomy Observatory (NRAO) is a
  facility of the National Science Foundation operated under
  cooperative agreement by Associated Universities, Inc.} VLBA and the
sample of stars consisted of the Mira variable stars U~Her, S~CrB,
RR~Aql, R~Aql, R~Cas and Y~Cas. Only for the first three of these
sources did we manage to detect maser emission that was strong enough
to image accurately. Their periods, velocities with respect to the
Local Standard of Rest (LSR), Hipparcos parallaxes and proper motions
and P--L distances are given in Table~\ref{sam1}. The observations of
all sources were scheduled in the same observing block for a total of
12~hours. This resulted in approximately 2~hours of observations per
maser source, including the time spent on the extragalactic phase
reference sources. Due to some observational problems (strong
interference, telescope and recording failures), a few hours of
observations at 3 of the epochs were unusable. Fortunately, none of
the three detected sources (U~Her, S~CrB and RR~Aql) were affected
more than once and we obtained 5 epochs of good data for each of
them. The resulting average beam size for the observations was
$14\times8$~mas.

\subsection{In-beam calibration}

To improve upon the results obtained in V03, we used in-beam
calibration in addition to the regular nodding phase referencing. As
shown in \citet{Chatterjee05}, the astrometric accuracy that can be
obtained by phase referencing is strongly dependent on the separation
between the target and reference source. After initial calibration on
a primary (nodding) phase reference source, residual calibration
errors arise primarily from the unmodeled ionosphere between target
and reference source. These errors can be significantly reduced by
using an secondary calibrator in the primary telescope beam, reducing
the angular throw as well as the need for time extrapolation
\citep[e.g.][]{Fomalont99}. Such a secondary in-beam calibrator can be
much fainter than the primary calibrator, with a flux $\gs 10$~mJy. In
an effort to find in-beam calibrators with sufficient flux that
additionally are suitably compact at VLBA baselines, we observed the
brightest NVSS sources \citep{Condon98} near a number of maser stars
with the Very Large Array (VLA) at 8.4~GHz. In addition to the stars
making up the sample presented here, we also observed possible
calibrator sources near W~Hya and RT~Oph. In total 40~sources were
observed for 3~min each. These were then reduced using AIPS without
any special processing. The images of all the sources, including those
that were not detected, are presented in
Figs.~\ref{calfig1},~\ref{calfig2},~\ref{calfig3} and~\ref{calfig4} in
the Online material. There we also present Table.~\ref{vlatab} with
positions and fluxes extracted using the AIPS {\it JMFIT} task. From
the VLA observations we determined the best VLBA in-beam
candidates. No good candidates were found for W~Hya and RT~Oph and
also for U~Her there was no obvious good in-beam candidate. However,
as U~Her is the longest astrometrically monitored OH maser source in
was still included in our sample even though in-beam calibration
turned out to be impossible. As in V03, we used J1630+2131 with a
separation of $2.8^\circ$ as primary calibrator for U~Her. For S~CrB,
the strongest in-beam calibrator source was also the primary
calibrator (J1522+3144) used in V03 at a distance of $24'$, so we were
able to accurately calibrate the masers of S~CrB without nodding the
telescopes. For RR~Aql we used both a primary calibrator (J2015-0137
at $4.4^\circ$ from RR~Aql) and an in-beam calibrator (NVSS
J195655-013615), with a target-calibrator separation of $\sim20'$ (see
Table~\ref{vlatab} in the on-line material for detailed coordinates).

\subsection{Data correlation}

We used a mixed bandwidth setting to observe both the 1667~MHz OH
maser line and the continuum calibrator simultaneously. One frequency
band with a bandwidth of $500$~kHz was centered on the stellar
velocity and correlated with moderate spectral resolution
($1.95$~kHz$~=0.36$ km~s$^{-1}$). Simultaneously three 4~MHz wide
bands were recorded to detect the continuum reference sources, with
one of the bands also covering the 1665~MHz OH line. These bands were
correlated twice with a spectral resolution of $15.6$~kHz, equaling
$2.88$~\kms. The first of the wide band correlator passes used the
position of the in-beam reference source for calibration purposes,
while the second pass used the stellar position to allow us to detect
the 1665~MHz OH maser line.

\subsection{Data calibration}

The astrometric VLBA data was calibrated using AIPS, with amplitude
calibration based on the antenna system temperatures and phase, rate
and delay calibration based on the primary and in-beam calibrators. We
applied ionospheric corrections using the AIPS task TECOR, but found
that these do not significantly improve our data
quality. Additionally, we used CLCOR to correct for errors in the
Earth Orientation Parameters (EOP) that affected VLBA observations
between May 2003 and August 2005 (Walker et al. 2005). For a
consistent solution we determined, when possible a model of the
primary and available in-beam calibrators using all epochs of data. As
the primary calibrator (J1630+2131) of U~Her showed structural changes
between the different epochs, this forced us to construct independent
calibrator models for every epoch. The models were then used to
iteratively calibrate the data.

As the calibration solutions from both the primary and the in-beam
calibrators are determined on the wide-band ($3\times4$~MHz) data,
special care needed to be taken connecting these solutions to the
narrow-band spectral line data. A special task written in AIPS for the
previous observations presented in vL00 and V03 was rewritten in the
ParselTongue scripting language \citep{Kettenis06} and was then used
to transfer the calibration solutions.

\subsection{Error analysis}
\label{error}

For the two sources with in-beam calibrators (S~CrB and RR~Aql), the
astrometric errors on the observational data points were taken to be
the uncertainty in calibrator and maser feature position added in
quadrature. The uncertainty in calibrator position is determined using
${\rm Beam}/[2\times{\rm SNR}_c]$ for unresolved calibrator
  sources , while that of the maser feature and of the resolved
  calibrator sources is $\theta_{c,m}/[2\times{\rm SNR}_{c,m}]$ with
${\rm SNR}_{c,m}$ the signal-to-noise ratio on the calibrator and
maser and $\theta_{c,m}$ the full width half-maximum size of the
  calibrator and the maser feature respectively. This size is equal
to the Beam-size when the maser is unresolved, which is typically the
case. The total positional error is of the order of $0.2$~mas during
the best observing conditions, with the error as a result of the
uncertainty in calibrator position typically smallest by an order of
magnitude. To total error increases to $\sim0.7$~mas for the weakest
maser-calibrator combinations and the worst ionospheric
conditions. The here described error estimates produced least
$\chi^2$-fits with reduced $\chi^2=1.01$ and $1.86$ for 5 degrees of
freedom for RR~Aql and S~CrB respectively.

For U~Her, where no suitable in-beam calibrator was found, astrometric
errors are expected to be larger. Fortunately, we were able to detect
both 1665 and 1667~MHz OH masers. As was found in vL00, the most
blue-shifted maser features in both transititions are coincident, and
correspond to the amplified stellar image. We thus estimate an
additional factor in the positional uncertainty from the internal
scatter between the 1665 and 1667~MHz most blue-shifted maser
features. This additional scatter factor, of the order of $\sim0.6$~mas, was also added quadratically to the positional error
estimate described above, producing a typical error of $\sim0.7$~mas. As a result, the $\chi^2$-fit to both 1667 and 1665~MHz OH
masers of U~Her simultaneously, has a reduced $\chi^2=1.61$ for 13
degrees of freedom.  These $\chi^2$ results imply that the error
estimates for all three of the sources are a fairly good
representation of the true astrometric errors.

Finally, we have checked the robustness of the astrometric
  results by randomly removing individual data points and performing
  new fits on the reduced data set. We find that for S~CrB and RR~Aql
  the fit solutions are consistent within $1\sigma$. For U~Her the
  solutions are consistent within $\sim 1.2\sigma$, hinting at a
  slight underestimate of the fitting errors for U~Her.

\section{Results}
\label{res}

\begin{figure}
   \resizebox{\hsize}{!}{\includegraphics{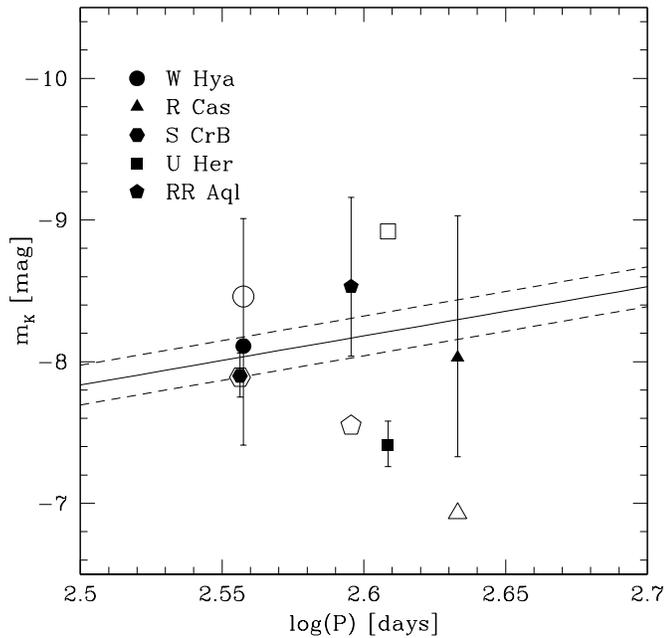}}
   \hfill \caption{Period vs. K$_0$ magnitudes for the OH maser stars with
     VLBI interferometric distances, including those from V03. The
     solid symbols and the error bars are determined using the VLBI
     distances. The open symbols are the values using the Hipparcos
     distances. The solid line is the P--L relation determined by
     \cite{Whitelock00} on the oxygen rich Mira stars observed
     with Hipparcos, the dashed lines are the spread in the relation
     due to the error in the P--L relation zero-point.}  
   \label{fig5.pl}
\end{figure}

The results of the $\chi^2$-fits are presented in
Table~\ref{res1}. The table contains the parallax and proper motion
results derived from the recent observational epochs as well as, for
comparison, the results from V03 for U~Her and S~CrB. For these
sources we have also performed a fit combining the entire data-set,
allowing for an additional zero-position off-set between the earlier
epochs and the data presented here. Such an additional off-set is
needed because of changes in the calibrator situation and since,
especially for S~CrB, we might be tracing a new maser
feature. Additionally, the errors for the early epochs in the combined
fitting have been increased by a factor of two, as in V03 only the
formal position fitting errors were used while in the original scheme
the ionospheric contributions could not be easily estimated easily
estimated. The table gives the velocity $V_{\rm LSR}$ of the maser
feature used for the fit with respect to the Local Standard of Rest
(LSR) as well as which maser transition (1665 or 1667~MHz) was used
for the fit.

\subsection{RR~Aql}

\begin{table*}
\caption{Results}
\begin{tabular}{|l|c|c||c|rr|c|c|c|}
\hline
Source & Maser & Note$^a$ & VLBI parallax & \multicolumn{2}{c|}{VLBI Proper motion} & \# epochs & $V_{\rm LSR}~^b$ & $d_{\rm VLBI}$\\
& (MHz) & & (mas) & \multicolumn{2}{c|} {RA, dec (mas/yr)} & & (\kms) & (pc)\\
\hline
\hline
\hline
S~CrB & 1667 & & $2.39 \pm 0.17$ & $-8.58 \pm 0.38$ & $-13.21 \pm 0.61$ & 5 & $4.6$~\kms & $418^{+21}_{-18}$\\
      & both & A & $2.31 \pm 0.33$ & $-9.08 \pm 0.27$ & $-12.49 \pm 0.33$ & 8 & $3.2/2.9$~\kms &\\
      & both & B & $2.36 \pm 0.23$ & $-9.06 \pm 0.23$ & $-12.52 \pm 0.29$ & 5+8 & &\\
\hline
U~Her & both & & $3.76 \pm 0.27$ & $-16.99 \pm 0.77$ & $-11.88 \pm 0.50$ & 5 & $-20.2$~\kms & $266^{+32}_{-28}$ \\
      & 1667 & A & $3.61 \pm 1.04$ & $-14.94 \pm 0.38$ & $-9.17 \pm 0.42$ & 12 & $-20.4$~\kms&\\
      & both & B & $3.74 \pm 0.61$ & $-14.98 \pm 0.29$ & $-9.23 \pm 0.32$ & 5+12 & &\\
\hline
RR~Aql & 1665 & & $1.58 \pm 0.40$ & $-25.11 \pm 0.74$ & $-49.82 \pm 0.54$ & 5 & $24$~\kms & $633^{+214}_{-128}$\\
\hline
\multicolumn{9}{l}{$^a$ A: results from V03; B: results from combined fit including V03 epochs}\\
\multicolumn{9}{l}{$^b$ LSR Velocity of fitted maser features}\\
\end{tabular}
\label{res1}
\end{table*}

RR~Aql is one of the new sources added to the sample, for which
Hipparcos did not manage to obtain a significant parallax. We did not
manage to detect any 1667~MHz maser emission, but were fortunately
able to image a $\sim0.1$~Jy/beam 1665~MHz maser feature in the low
spectral resolution data. The feature was found at $24$~\kms,
blue-shifted with respect to the stellar velocity, but at a velocity
unlikely to belong to an amplified stellar image. We managed to detect
and image the feature at 5 epochs with respect to the nearby in-beam
calibrator. This yielded the smallest parallax determined to date for
OH maser astrometry, even though the errors are dominated by the fact
that the maser is weak and is only detected in the low resolution
spectral line data-set.

\subsection{S~CrB}

S~CrB was previously monitored with the VLBA (V03), at which time we
were able to detect bright masers at both 1665 and 1667~MHz (at
$V_{\rm LSR}=3.2$ and $2.9$~\kms~respectively). In the new epochs we
were only able to detect a 1667~MHz maser feature, red-shifted with
respect to the stellar velocity, at $V_{\rm LSR}=4.6$~\kms. The
feature was fairly weak during the first 2 epochs, with the flux
decreasing from $0.07$~Jy/beam in the first to below $0.05$~Jy/beam in
the second epoch. The third epoch turned out to be unusable, but
during the last three epochs the maser had brightened again to $\sim0.11$~Jy/beam. Thus, the astrometric errors were almost a factor of
two smaller in the last three epochs. Both the parallax and proper
motion fit results from the newest epochs are fully consistent with
the result obtained from the earlier epochs.

\subsection{U~Her}

U~Her is the OH maser source which has been the target of the longest
astrometric VLBI monitoring campaign, and currently has data spanning
almost 12 years (vL00, V03). In the new observations, we again
detected the most blue-shifted maser feature corresponding to the
amplified stellar image in both the 1665 and 1667~MHz maser
transition, with a flux varying between $0.8$ and $1.1$~Jy/beam. In
addition, we detected a number of weaker features which are discussed
in \S~\ref{uher}. While the parallax is also fully consistent with the
earlier determined parallax, albeit with a significantly smaller
error, the proper motion has changed considerably. As was already
found in V03, the calibrator used for phase referencing, J1630+2131,
shows structural changes on timescales at the order of several
months. This is possibly the cause of the differences in proper
motions. Fig~\ref{oJ1630} of the online material shows a series of
images of J1630+2131 for each observational epoch indicating these
changes. As described in V03, the internal calibrator motions are
  expected to be relatively slow compared to the annual timescale of
  the parallax and we thus think it unlikely that this will bias the
  parallax measurement. However, the motions will introduce additional
  scatter and some caution has to be taken with respect to the quoted
  uncertainties.  Because of the proper motion difference, caution
has to be taken in using the fitted motion resulting from the combined
epochs fitting.

\section{Discussion}
\label{disc}

\subsection{Comparisson with previous results}

The distances to the OH bearing Mira variable stars S~CrB and U~Her
derived in this paper are fully consistent with the previously
published values in V03. The uncertainty on the parallax measurement
decreased by over a factor of 2, even with significantly fewer
observational epochs, illustrating the astrometric improvements. The
errors on the proper motion measurements are larger, mainly due to the
shorter time-baseline of the new observations. The most accurate
proper motions have therefore been obtained by simultaneously fitting
the maser positions from both the current and previous observational
epochs as described above.
 
The astrometric measurements are also a significant improvement on the
Hipparcos derived values, which suffered from the fact that the stars
in our sample are relatively faint, variable, AGB stars (see V03 for
more details). When comparing the VLBI derived distances in
Table~\ref{res1} with distances derived from a P--L relation in
Table~\ref{sam1}, we find that, while for both S~CrB and RR~Aql the
distances are reasonably consistent, the U~Her VLBI distance is
significantly lower. This can also be seen in Fig.~\ref{fig5.pl},
which shows position of the OH maser stars with VLBI parallaxes in the
P--L diagram. As was speculated in vL00 and V03, the discrepancy for
U~Her could be a result from U~Her pulsating in the fundamental mode,
while the other stars are pulsating in the first overtone
mode. However, as it has been argued that all Mira stars are pulsating
in the fundamental mode \citep{Wood98}, the offset of U~Her more likely
highlights the scatter in the P--L relation or the uncertainty in
individual magnitude determinations. We rule out that the offset is introduced by the absence of an in-beam calibrator for this source, because we can imagine nu systematic effect that can bias the parallax to larger values over all the years we have monitored the source.

\subsection{Morphology and internal motions of the OH masers of U~Her}
\label{uher}

\begin{figure*}
   \resizebox{\hsize}{!}{\includegraphics{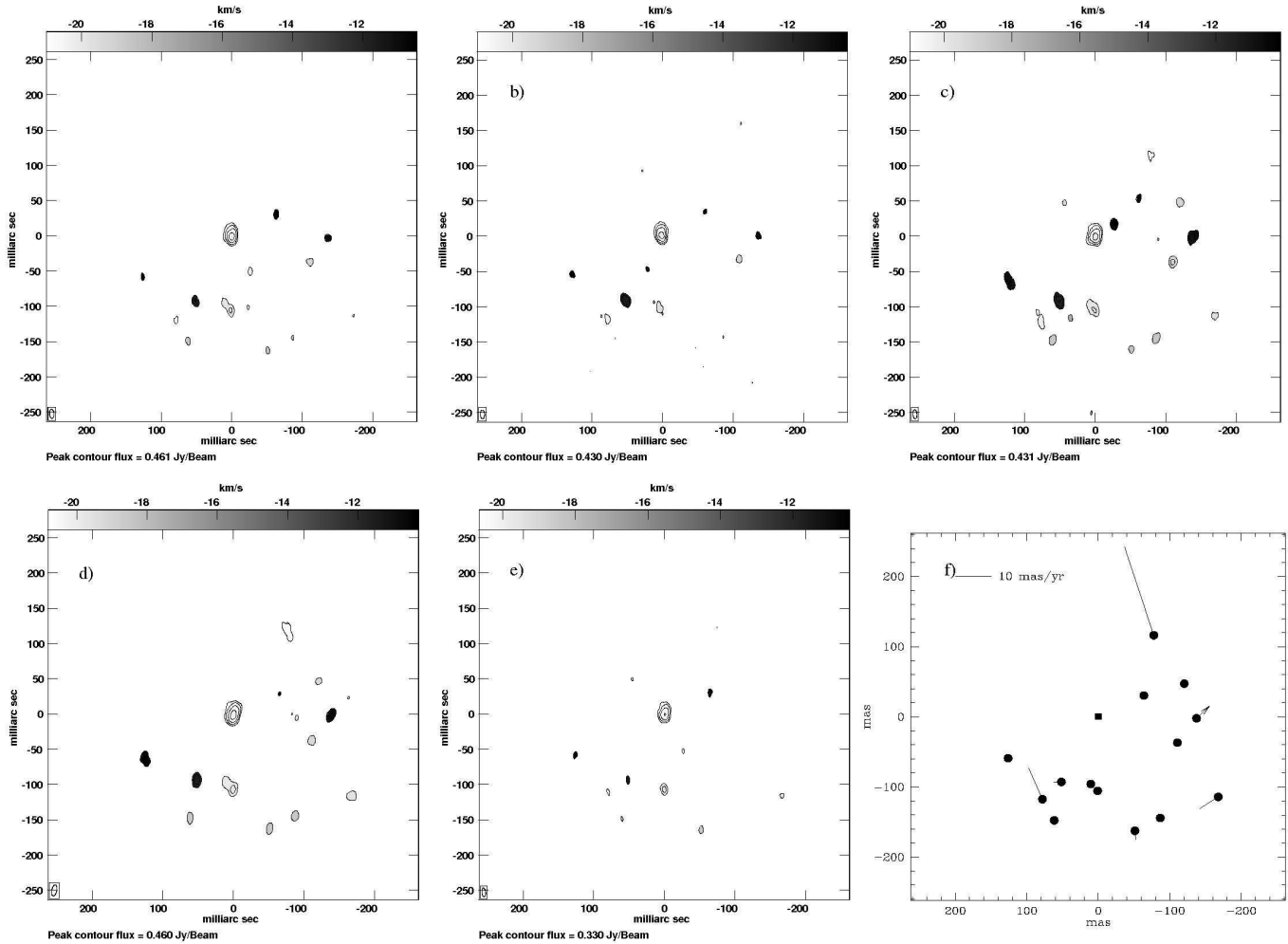}}
   \hfill \caption{(a-e) Images for 5 epochs of the 1667~MHz OH maser
     emission of U~Her referenced to the most brightest maser. The
     brightest and most blue-shifted maser feature is prominent in all
     images and is identified with the stellar image amplified by the
     masing material at the front side of the expanding shell
     (V03). The greyscale indicates the first moment map with the
     velocity scale indicated at the top of the figures. The contours
     indicate the total intensity I, drawn at levels of 40, 80, 160
     and 320~mJy/Beam. The rms noise varies somewhat for the different
     epochs but is $\sim12$~mJy/Beam. The peak intensity is given for
     each individual image. (f) The maser features that were observed
     during at least 2 epochs with vectors indicating their proper
     motion. The arrow denotes a $3\sigma$ proper motion
     detection, no significant proper motion was detected for any of
     the other features but for illustration purposes we show
     $1$--$3\sigma$ proper motions with the regular vectors. The solid
     square indicates the position of the stellar image.}
   \label{fig4.uher}
\end{figure*}

While for S~CrB and RR~Aql we only detect maser features in a few
neighbouring red-shifted and blue-shifted spectral channels
respectively, we observe a large number of both blue- and red-shifted
features for U~Her. Fig.~\ref{fig4.uher}(a-e) show the first moment
greyscale maps with overplotted total intensity contours for the 5
succesful U~Her epochs. The maps are all referenced the most blue-shifted maser feature that has been identified as the stellar
image. Several of the maser features are detected in multiple epochs
and these are shown in Fig.~\ref{fig4.uher}(f). The brightest
red-shifted masers are all detected with $V_{\rm LSR}$ between $-10$
and $-12$~\kms while the blue-shifted masers are located at $V_{\rm
  LSR}$ between $-17$ and $-21$~\kms. There is no obvious difference
in the extent of the red- and blue-shifted maser features, which are
both spread over $\sim200\times200$~mas, corresponding to $\sim55\times55$~AU at the $266$~pc distance of U~Her. Fig.~\ref{fig4.uher} does
show that the maser distribution is distinctly asymmetrical with
respect to the stellar image, with hardly any maser features detected
to the North-East. This is significantly different from the 1665 and
1667~GHz maser distribution observed in 1984 with MERLIN
\citep{Chapman94}, but similar to our previous 1665 and 1667~GHZ OH
maser observations between 1995 and 2002 (vL00, V03). As the 22~GHz
H$_2$O maser emission is also clearly weakest towards the North-East
\citep{Vlemmings02, Bains03}, the observed non-spherical maser
distribution is likely due to a real asymmetry in the CSE.

As several U~Her maser features, besides the stellar image, can be
traced over multiple epochs, we are able to study the internal motions
in the maser envelope. One red-shifted maser feature, at $V_{\rm
  LSR}=-11.51$~\kms, was persistent over the entire year of
observation, while three maser features are found in 4 epochs. A
further 5 features can be traced over three epochs while another 5 are
detected in only 2 epochs. Although none of these features are bright
enough to be included in the regular astrometric fitting, we are able
to test the consistency of our observations by determining the motions
of these features after self-calibrating on the blue-shifted maser
feature amplifying the stellar emission. By applying the
self-calibration we remove the stellar proper motion and parallax,
after which we fit for residual proper motions of the maser
features. For most of the maser features we do not detect significant
proper motion. Only for one red-shifted maser feature, that was
detected in the first three epochs at $V_{\rm LSR}=-10.81$~\kms, we
detect a proper motion of $\mu = 5.1 \pm 1.5$~mas~yr$^{-1}$, which
corresponds to a velocity of $6.4 \pm 1.9$~\kms at the distance to
U~Her. Assuming a spherically expanding shell, this implies, from
  the angular off-set to the stellar image and the radial velocity, an
  outflow velocity of $\sim7.5$~\kms in the U~Her OH maser region.
The proper motion of this feature is indicated with a thick line in
Fig.~\ref{fig4.uher}. Performing a weighted ensemble average of the
masers in three quadrants ($\alpha<0$ and $\delta>0$, $\alpha<0$ and
$\delta<0$, $\alpha>0$ and $\delta<0$) reveals no significant
motions. Considering that we only observe the front and back side of
the maser shell this is not surprising, as the outflow will be
predominantly in the radial direction. Thus, as no residual proper
motion is detected, the motion determined from the astrometric fitting
to the amplified stellar image can be taken as the true stellar proper
motion, with only a possible contribution from the internal motions of
the phase reference source.

As we have concluded that the majority of the maser spots show no
significant proper motion when referencing them to the stellar image,
the residual motion of the maser features is likely due to a
combination of the intrinsic astrometric errors and turbulent motions
in the masing medium. We find that the rms scatter of the masers
detected at multiple epochs is $\sigma_{\rm pos} = 2.0 \pm
1.3$~mas. Taking into account the intrinsic positional fitting
uncertainty, including the error discussed in \S\ref{error}, which for
these weak maser features is of the order of $1.5$~mas, small but
significant residual motions are present . These we attribute to
turbulence at the level of $1-2$~\kms, only slightly higher than the
typical assumed value of $1$~\kms \citep{Diamond85}.

\section{Concluding remarks}
\label{concl}

We have been able to significantly improve the parallax measurements
for a number of OH maser stars. The improvements were the results of
using nearby phase reference sources in the form of in-beam
calibrators, and generally good observing conditions due to a minimum
in solar activity. As a result the parallax uncertainties were
decreased by up to a factor of $\sim4$ to the order of $\sim0.2$~mas,
using only 5 observing epochs compared to the $8$--$12$ epochs
presented in V03. Thus, we have convincingly shown the capability of
VLBI astrometry to determine distances up to 1~kpc to AGB stars using
their circumstellar OH masers, even when no special maser feature
amplifying the stellar image is found. The limit on distance
measurements could potentially be stretched to $\sim 2$~kpc or $\pi
\sim 0.5$~mas when the OH maser is strong ($\gs 0.25$~Jy/beam) and a
bright compact in-beam calibrator is available. The precision will be
further increased when a number of different maser features can be
traced over a sufficient number of epochs. This opens up the
possibility to use OH maser astrometry to determine the parallaxes to
a large number of OH maser stars with previously unknown distances,
which can for instance be used to contruct a P--L relation for the most
heavily enshrouded AGB stars.

{\it acknowledgments:} This research was partly supported by a Marie Curie
Intra-European fellowship within the 6th European Community Framework
Program under contract number MEIF-CT-2005-010393. WV thanks the
Max-Planck Institute for Astronomy in Heidelberg for the regular
hospitality. We acknowledge the effort of Daniele Biancu for
implementing the mixed bandwidth calibration scheme in
ParselTongue. ParselTongue is developed within the context of the
RadioNet Joint Research Activity ALBUS. This work has benefited from
research funding from the European Community's sixth Framework
Programme under RadioNet R113CT 2003 5058187.

%
%
 \Online

 \begin{figure*}
 \centering
 \resizebox{0.9\hsize}{!}{\includegraphics{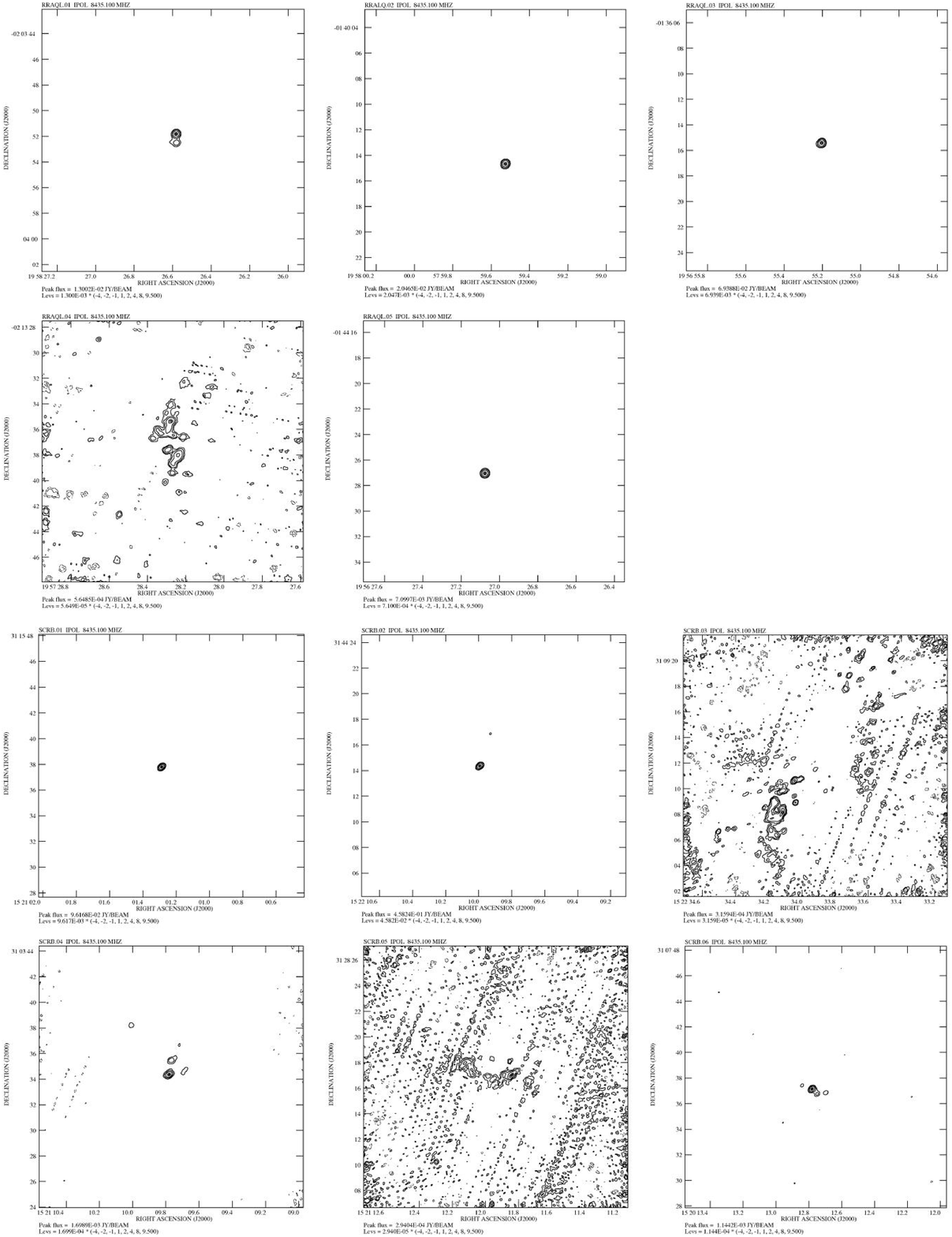}}
 \caption{Results from the 8.4~GHz VLA in-beam calibrator search for
   RR~Aql and S~CrB. The images are labelled according to
   Table~\ref{vlatab} and include non-detections for completeness. No
   special calibration to improve the images was performed. Peak flux
   and contour levels are given for each image individually. For
   RR~Aql and S~CrB, VLBA in-beam calibration was sucessfully
   performed on the sources RRAQL.03 and SCRB.02.}\label{calfig1}
 \end{figure*}

 \begin{figure*}
 \centering
 \resizebox{0.8\hsize}{!}{\includegraphics{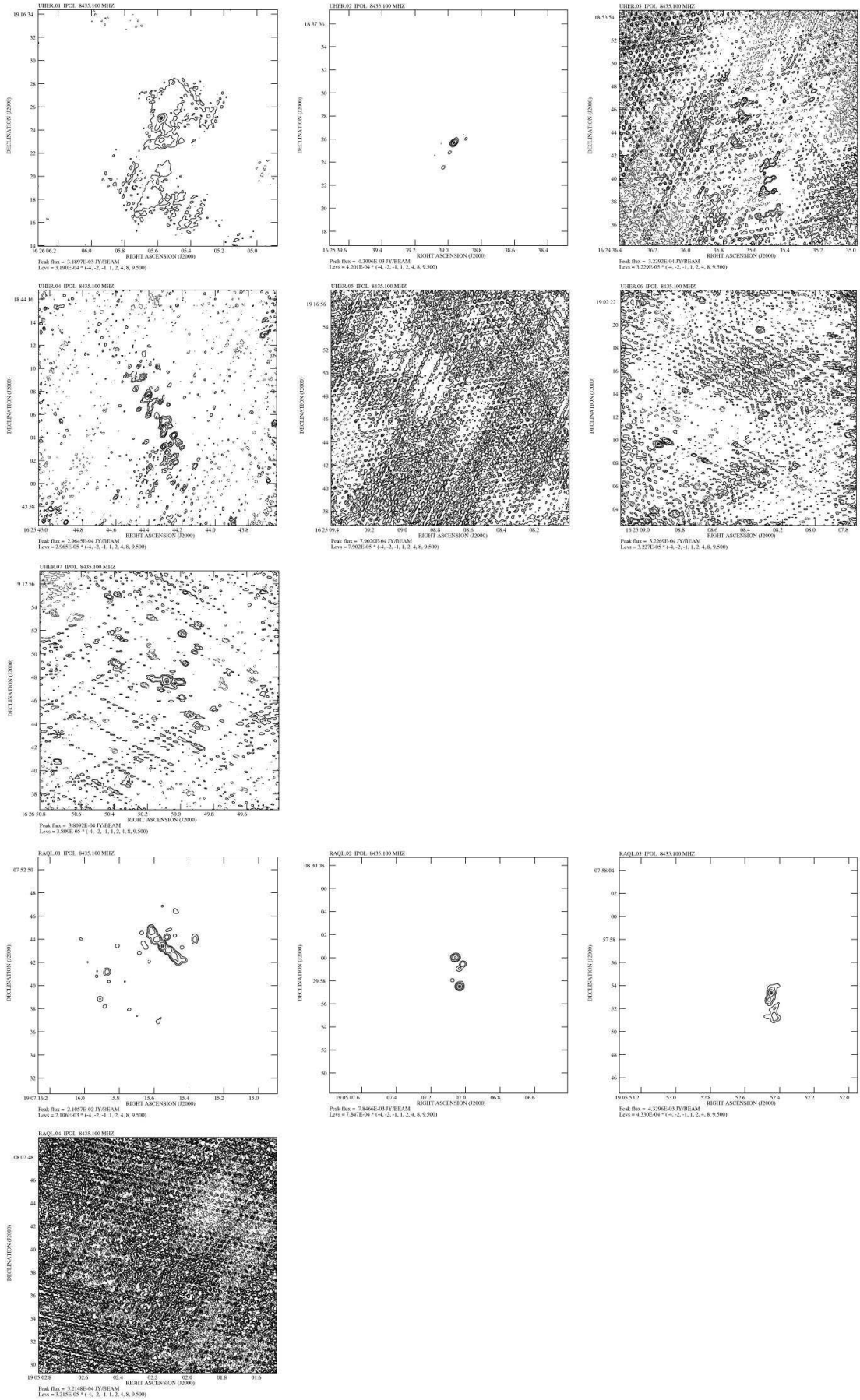}}
 \caption{As Fig.~\ref{calfig1} for U~Her and R~Aql. VLBA in-beam
   calibration was unsuccesfully attempted on UHER.02. No masers were
   detected for R~Aql during the VLBA observations.}\label{calfig2}
 \end{figure*}

 \begin{figure*}
 \centering
 \resizebox{0.9\hsize}{!}{\includegraphics{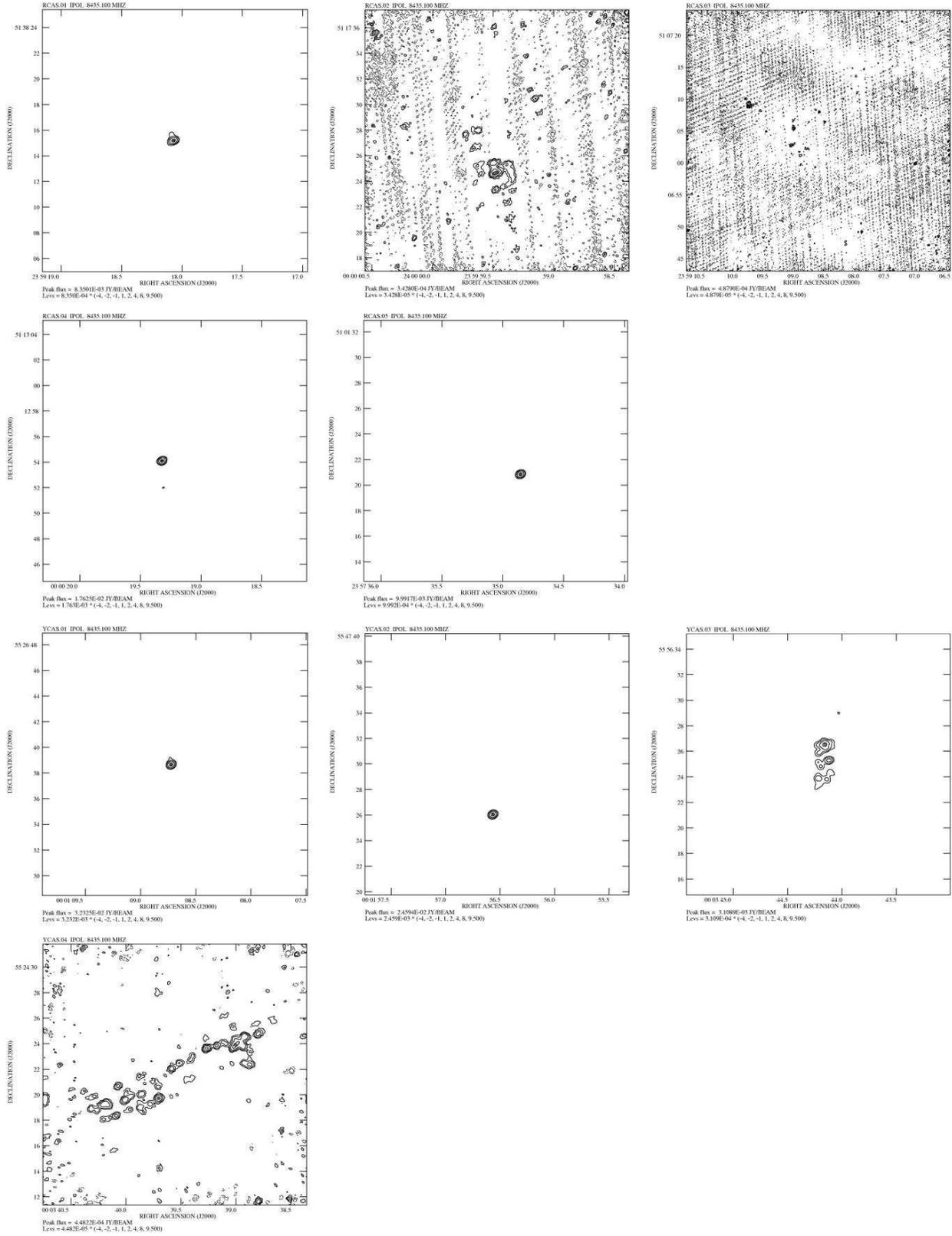}}
 \caption{As Fig.~\ref{calfig1} for R~Cas and Y~Cas. No masers were detected during the VLBA observations for either star.}\label{calfig3}
 \end{figure*}

 \begin{figure*}
 \centering
 \resizebox{0.9\hsize}{!}{\includegraphics{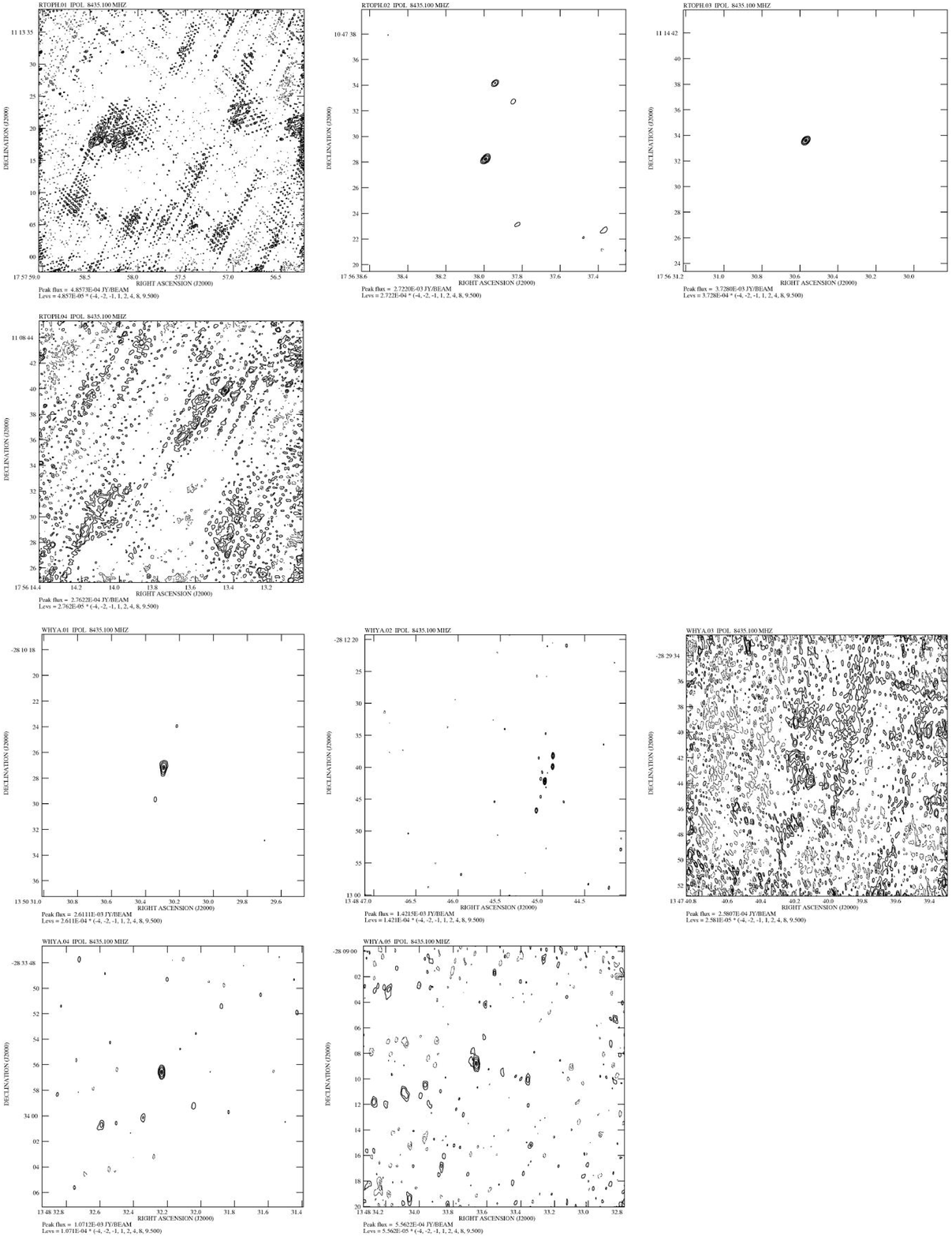}}
 \caption{As Fig.~\ref{calfig1} for RT~Oph and W~Hya. None of the
   sources were of sufficient quality to be used as in-beam
   calibrators.}\label{calfig4}
 \end{figure*}

\begin{table*}
\caption{X-Band VLA observation results: Results from the X-band VLA search for suitable in-beam
  calibrators. The sources were picked from the 1.4~GHz NVSS survey
  \citep{Condon98}. The observations were taken in D-configuration at May 21st 2003. The table gives the name identifying the nearby
  maser source with the NVSS name and integrated flux. Additionally,
  we give the coordinates of the peak emission, and the integrated and peak 8.4~GHz flux determined using the AIPS task {\it JMFIT}, when a single
  component Gaussian fit was possible for the strongest
  component. The last column contains the seperation $\Delta\theta$ between the continuum source and the maser star in
  arcminutes. For those sources that were used as in-beam calibrators and detected  in the VLBA L-band observations we also give {\it JMFIT} intergrated fluxes.}
\begin{tabular}{|l|l|c|c|c|c|c|c|}
\hline
{Source} & {NVSS name} & NVSS Flux & $\alpha_{\rm J2000}$ & $\delta_{\rm J2000}$ & Flux & Peak Flux & $\Delta\theta$ \\
 & & (mJy) & $hh~mm~ss$ & $^\circ~'~''$ & (mJy) & (mJy/beam) & (arcmin) \\
\hline
\hline
RRAQL.01 & NVSS J195826-020352 & $116.8\pm3.5$ & $19~58~26.5826$ & $-02~03~51.833$ & $13.57\pm0.19$ & $12.77\pm0.11$ & $16.5$ \\
RRAQL.02 & NVSS J195759-014012 & $115.5\pm4.1$ & $19~57~59.5262$ & $-01~40~14.670$ & $21.00\pm0.08$ & $20.48\pm0.05$ & $14.2$ \\
RRAQL.03$^a$ & NVSS J195655-013615 & $94.3\pm2.9$ & $19~56~55.2046$ & $-01~36~15.426$ & $70.96\pm0.12$ & $69.95\pm0.07$ & $19.8$ \\
 & L-band VLBI & $48.1\pm0.2$ & & & & & \\
RRAQL.04 & NVSS J195728-021337 & $36.6\pm1.2$ & & & & & $20.5$ \\
RRAQL.05 & NVSS J195627-014425 & $16.0\pm0.6$ & $19~56~27.0759$ & $-01~44~27.040$ & $7.20\pm0.04$ & $7.14\pm0.02$ & $19.3$ \\
\hline
SCRB.01 & NVSS J152101+311537 & $175.5\pm5.3$ & $15~21~01.2872$ & $+31~15~37.796$ & $97.48\pm0.69$ & $96.99\pm0.40$ & $8.1$ \\
SCRB.02$^{a,b}$ & NVSS J152209+314414 & $347.2\pm10.4$ & $15~22~09.9937$ & $+31~44~14.361$ & $465.18\pm2.95$ & $458.56\pm1.69$ & $24.2$ \\
 & L-band VLBI & $252.7\pm0.1$ & & & & & \\
SCRB.03 & NVSS J152233+310911 & $51.3\pm1.9$ & & & & & $19.7$ \\
SCRB.04 & NVSS J152109+310334 & $21.3\pm0.7$ & $15~21~09.7814$ & $+31~03~34.369$ & $2.44\pm0.11$ & $1.62\pm0.05$ & $18.7$ \\
SCRB.05 & NVSS J152012+310738 & $11.9\pm0.5$ & & & & & $20.9$ \\
SCRB.06 & NVSS J152111+312816 & $11.8\pm0.5$ & $15~20~12.7797$ & $+31~07~37.135$ & $1.27\pm0.03$ & $1.11\pm0.01$ & $6.8$ \\
\hline
UHER.01 & NVSS J162605+191624 & $114.0\pm3.4$ & $16~26~05.5769$ & $+19~16~24.999$ & $9.24\pm0.07$ & $2.63\pm0.02$ & $23.2$ \\
UHER.02$^a$ & NVSS J162539+183727 & $32.8\pm1.1$ & $16~25~38.9655$ & $+18~37~25.691$ & $5.10\pm0.02$ & $3.92\pm0.01$ & $16.2$ \\
 & L-band VLBI & $4.5\pm0.3$ & & & & & \\
UHER.03 & NVSS J162435+185344 & $29.9\pm1.0$ & $16~24~35.6826$ & $+18~53~45.245$ & $0.78\pm0.10$ & $0.29\pm0.03$ & $17.0$ \\
UHER.04 & NVSS J162544+184406 & $20.2\pm0.7$ & & & & & $9.5$ \\
UHER.05 & NVSS J162508+191647 & $20.0\pm1.1$ & & & & & $25.0$ \\
UHER.06 & NVSS J162508+190212 & $7.8\pm0.5$ & & & & & $12.7$ \\
UHER.07 & NVSS J162650+191246 & $12.8\pm0.6$ & $16~26~50.0838$ & $+16~26~50.0838$ & $0.94\pm0.06$ & $0.37\pm0.02$ & $24.3$ \\
\hline
RAQL.01$^a$ & NVSS J190715+075240 & $268.1\pm9.5$ & $19~07~15.5506$ & $+07~52~43.409$ & $35.07\pm0.28$ & $20.10\pm0.11$ & $24.9$ \\
 & L-band VLBI & $1.1\pm0.2$ & & & & & \\
RAQL.02 & NVSS J190507+082958 & $200.0\pm6.0$ & $19~05~07.0343$ & $+08~29~57.507$ & $8.95\pm0.07$ & $7.66\pm0.04$ & $24.6$ \\
RAQL.03 & NVSS J190552+075754 & $105.6\pm3.2$ & $19~05~52.4491$ & $+07~57~53.358$ & $6.88\pm0.05$ & $3.80\pm0.02$ & $17.5$ \\
RAQL.04 & NVSS J190502+080239 & $56.0\pm3.5$ & & & & & $22.7$ \\
\hline
RCAS.01 & NVSS J235918+513815 & $58.9\pm1.8$ & $23~59~19.0708$ & $+51~38~15.198$ & $9.97\pm0.07$ & $8.11\pm0.03$ & $17.1$ \\
RCAS.02 & NVSS J235959+511727 & $8.7\pm0.5$ & & & & & $15.9$ \\
RCAS.03 & NVSS J235908+510703 & $7.3\pm0.5$ & $23~59~09.8053$ & $+51~07~08.929$ & $0.50\pm0.04$ & $0.47\pm0.02$ & $17.7$ \\
RCAS.04$^a$ & NVSS J000019+511254 & $6.9\pm0.5$ & $00~00~19.3222$ & $+51~12~54.106$ & $17.49\pm0.17$ & $17.60\pm0.10$ & $20.7$ \\
RCAS.05 & NVSS J235735+510122 & $6.8\pm0.5$ & $23~57~34.8546$ & $+51~01~20.852$ & $10.03\pm0.06$ & $10.06\pm0.03$ & $23.3$ \\
\hline
YCAS.01$^a$ & NVSS J000108+552638 & $289.7\pm8.7$ & $00~01~08.7229$ & $+55~26~38.668$ & $33.93\pm0.11$ & $31.76\pm0.06$ & $23.5$ \\
 & L-band VLBI & $68.5\pm0.3$ & & & & & \\
YCAS.02 & NVSS J000156+554730 & $115.8\pm4.2$ & $00~01~56.5666$ & $+55~47~26.041$ & $24.82\pm0.10$ & $24.56\pm0.06$ & $13.7$ \\
YCAS.03 & NVSS J000344+555625 & $84.4\pm2.6$ & $00~03~44.1589$ & $+55~56~26.483$ & $7.80\pm0.07$ & $2.68\pm0.02$ & $15.9$ \\
YCAS.04 & NVSS J000339+552421 & $42.6\pm1.7$ & & & & & $16.7$ \\
\hline
RTOPH.01 & NVSS J175757+111318 & $58.8\pm2.1$ & & & & & $21.1$ \\
RTOPH.02 & NVSS J175637+104729 & $37.7\pm1.2$ & $17~56~37.9921$ & $+10~47~28.239$ & $3.15\pm0.05$ & $2.70\pm0.03$ & $22.7$ \\
RTOPH.03 & NVSS J175630+111433 & $22.8\pm0.8$ & $17~56~30.5689$ & $+11~14~33.598$ & $3.77\pm0.03$ & $3.70\pm0.02$ & $4.4$ \\
RTOPH.04 & NVSS J175613+110835 & $16.1\pm0.6$ & & & & & $4.9$ \\
\hline
WHYA.01 & NVSS J135030-281027 & $16.3\pm0.7$ & $13~50~30.2936$ & $-28~10~27.173$ &$2.97\pm0.09$ & $2.56\pm0.05$ & $22.6$ \\
WHYA.02 & NVSS J134845-281239 & $9.9\pm0.6$ & $13~48~44.9729$ & $-28~12~42.180$ & $1.49\pm0.05$ & $1.41\pm0.03$ & $10.1$ \\
WHYA.03 & NVSS J134740-282942 & $8.4\pm0.5$ & & & & & $19.6$ \\
WHYA.04 & NVSS J134832-283356 & $7.7\pm0.5$ & $13~48~32.2336$ & $-28~33~56.579$ & $1.08\pm0.03$ & $1.08\pm0.02$ & $13.6$ \\
WHYA.05 & NVSS J134833-280909 & $7.3\pm0.5$ & $13~48~33.6641$ & $-28~09~08.811$ & $0.67\pm0.04$ & $0.54\pm0.02$ & $14.3$ \\
\hline
\multicolumn{8}{l}{$^a$ In-beam calibrator used during VLBA observations.}\\
\multicolumn{8}{l}{$^b$ Also known as JVAS source J1522+3144 with previously know VLBI position.}\\
\end{tabular}
\label{vlatab}
\end{table*}

 \begin{figure*}
 \resizebox{\hsize}{!}{\includegraphics{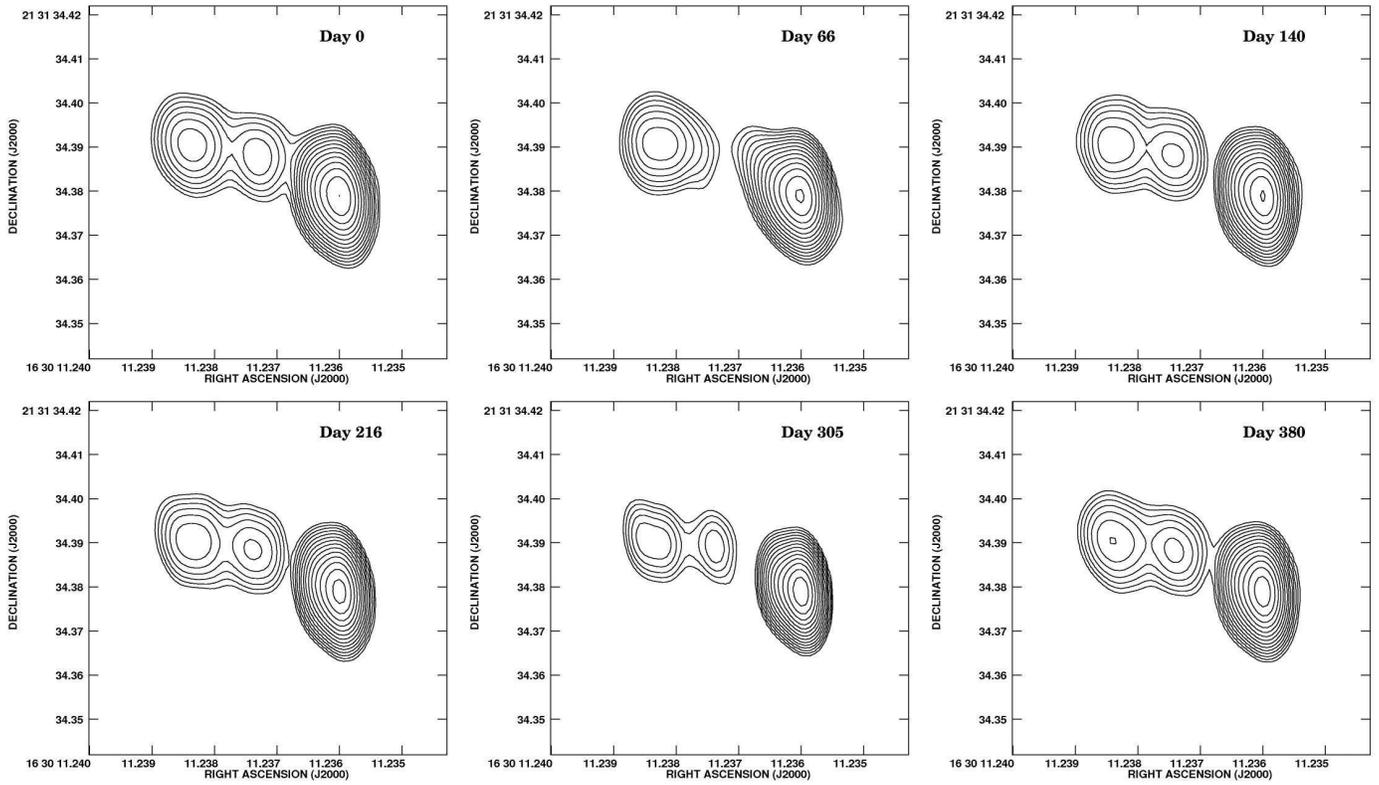}}
 \caption{The primary phase reference calibrator J1630+2131 used for U~Her. The
   panels are labelled by the number of days after the first epoch of
   observations and include epoch $e$, for which we were unable to
   properly image the OH masers. The contours start at $1$~mJy and
   each subsequent contour level is multiplied by a factor of
   $\sqrt{2}$.} \label{oJ1630}
 \end{figure*}

\end{document}